\documentclass[%
 reprint,
 amsmath,amssymb,
 aps,
floatfix,
]{revtex4-2}

\usepackage{esint}

\makeatletter
\@for\next:={int,iint,iiint,iiiint,dotsint,oint,oiint,sqint,sqiint,
  ointctrclockwise,ointclockwise,varointclockwise,varointctrclockwise,
  fint,varoiint,landupint,landdownint}\do{%
    \expandafter\edef\csname\next\endcsname{%
      \noexpand\DOTSI
      \expandafter\noexpand\csname\next op\endcsname
      \noexpand\ilimits@
    }%
  }
\makeatother

\usepackage{graphicx}% Include figure files
\usepackage[percent]{overpic}
\usepackage{dcolumn}% Align table columns on decimal point
\usepackage{bm}% bold math
\usepackage[normalem]{ulem}% for text fonts
\usepackage{float}%for figure placement
\usepackage[bottom]{footmisc}
\usepackage{xcolor}
\usepackage{physics}
\usepackage{tikz}
\usetikzlibrary{shapes.geometric}
\usetikzlibrary{calc}
\usepackage{hyperref}% add hypertext capabilities
\hypersetup{colorlinks=true, linkcolor=violet, citecolor=blue}
\begin{document}

\preprint{APS/123-QED}

\title{Will a single two-level atom simultaneously scatter two photons?}

\author{Luke Masters}
\author{Xinxin Hu}
\author{Martin Cordier}
\author{Gabriele Maron}
\author{Lucas Pache}
\author{Arno Rauschenbeutel}
\author{Max Schemmer}
\author{J\"urgen Volz}
    \email{juergen.volz@hu-berlin.de}

\affiliation{Department of Physics, Humboldt Universit\"at zu Berlin, 10099 Berlin, Germany}

%\date{\today}
\date{September 2, 2022}

\begin{abstract}
The interaction of light with a single two-level emitter is the most fundamental process in quantum optics, and is key to many quantum applications \cite{walmsey15, gisin07, ladd10, kimble08}. As a distinctive feature, two photons are never detected simultaneously in the light scattered by the emitter \cite{carmichael76, kimble77}. This is commonly interpreted by saying that a single two-level quantum emitter can only absorb and emit single photons. However, it has been theoretically proposed \cite{dalibard83} that the photon anti-correlations can be thought to arise from quantum interference between two possible two-photon scattering amplitudes, which one refers to as coherent and incoherent. This picture is in stark contrast to the aforementioned one, in that it assumes that the atom even has two different mechanisms at its disposal to scatter two photons at the same time. Here, we validate the interference picture by experimentally verifying the 40-year-old conjecture that, by spectrally rejecting only the coherent component of the fluorescence light of a single two-level atom, the remaining light consists of photon pairs that have been simultaneously scattered by the atom. Our results offer fundamental insights into the quantum-mechanical interaction between light and matter and open up novel approaches for the generation of highly non-classical light fields.
\end{abstract}

\maketitle

\par The interaction of a single two-level quantum emitter with a near-resonant coherent light field is one of the cornerstones of quantum optics and lies at the heart of many modern experiments and applications in this field \cite{eisaman11, aharonovich16}. The quantum mechanical description of this interaction \cite{GerryKnight04, Steck07} shows that the scattered light exhibits photon antibunching, i.e., it never contains two photons at the same time and place. This property can be seen as a consequence of the photon emission being associated with quantum jumps from the emitter's excited state to its ground state.
A more in-depth inspection of the optical Bloch equations reveals the existence of two distinct components of the scattered field, referred to as coherently and incoherently scattered light, reflecting their respective ability and inability to interfere with the driving field. Interestingly, when considered individually, these two components also contain higher photon-number components, i.e. two or more photons at the same time and place. In view of this fact, it has been argued that the origin of antibunching in resonance fluorescence stems from the destructive interference between the higher photon-number components of the coherently and incoherently scattered light \cite{dalibard83}. In this context, it has been recently demonstrated that when spectrally rejecting the incoherently scattered component, all photon correlations are lost, and the remaining light features the same characteristics as the classical driving field \cite{hanschke20,phillips20}. Intriguingly however, almost 40 years ago, it was conjectured that when performing the opposite, i.e. rejecting the coherent component, the remaining incoherent component consists of maximally correlated photons that have been scattered simultaneously by the emitter.

\begin{figure}[b]
\centering
    \includegraphics[width=0.48\textwidth,keepaspectratio]{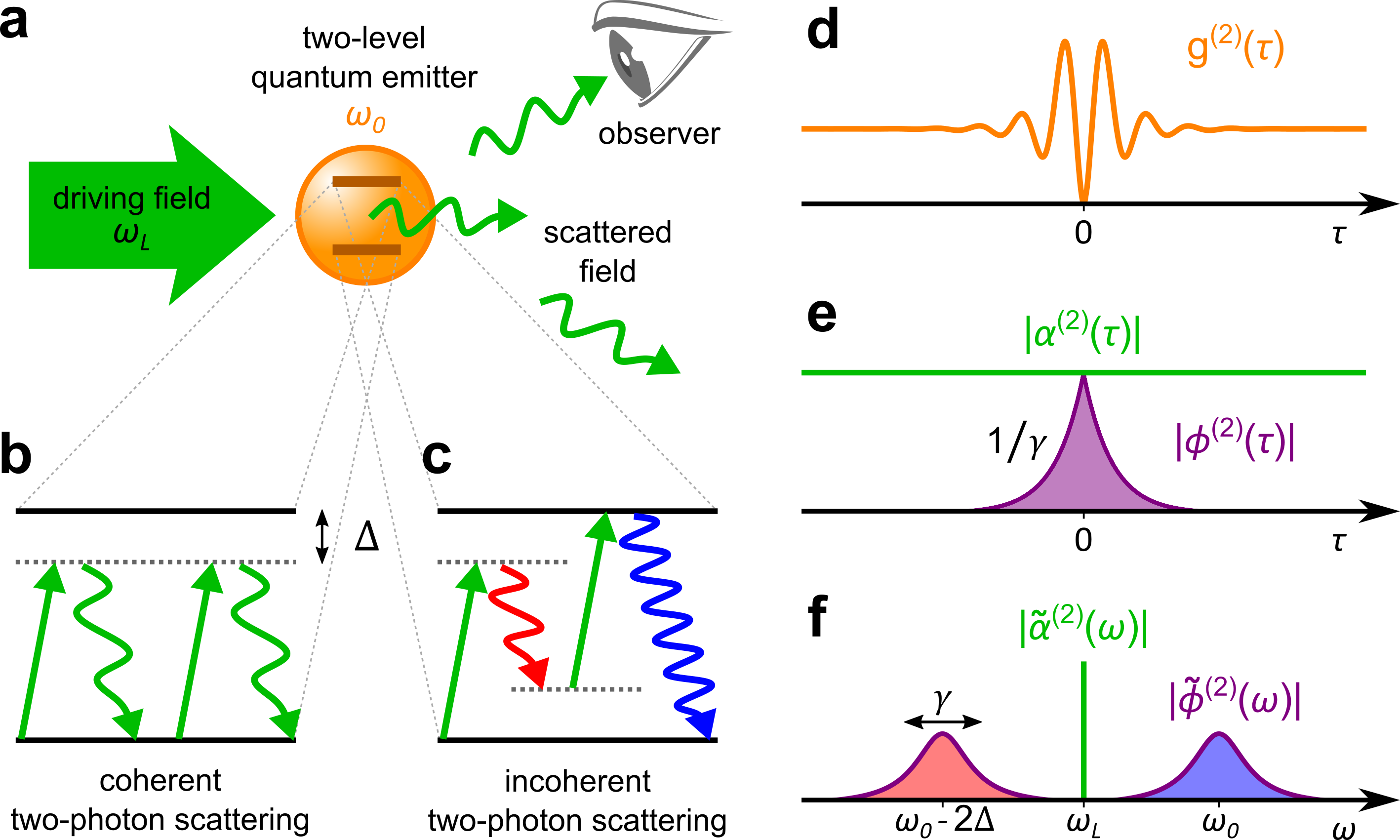}
\caption{\label{Experiment_concept} \textbf{Two-photon scattering processes in resonance fluorescence}. (a) Observation of resonance fluorescence from a two-level quantum emitter of resonance frequency $\omega_0$, driven by a light field at frequency $\omega_L = \omega_0 + \Delta$, where $\Delta$ denotes the detuning. (b) Coherent two-photon Rayleigh scattering leads to a pair of uncorrelated photons. (c) Incoherent two-photon scattering leads to an energy-time entangled pair of photons that are, in general, detuned from the driving field.
(d) For zero time delay $\tau$, the theoretically predicted second order correlation function yields $g^{(2)}(\tau = 0) = 0$. For $\lvert\tau\rvert > 0$, Rabi oscillations are apparent.
(e) Temporal two-photon wavefunctions of the coherent and incoherent component, $\alpha^{(2)}(\tau)$ and $\phi^{(2)}(\tau)$ respectively, where $\phi^{(2)}(0) = -\alpha^{(2)}(\tau)$. (f) Two-photon wavefunctions in the frequency domain. The coherent component, $\tilde{\alpha}^{(2)}(\omega)$, is a delta-function centered at $\omega_L$, while the incoherent component, $\tilde{\phi}^{(2)}(\omega)$, is given by a double Lorentzian at $\omega_L \pm \Delta$.}
\end{figure}

\begin{figure*}[t]
\centering
    \includegraphics[width=\textwidth,keepaspectratio]{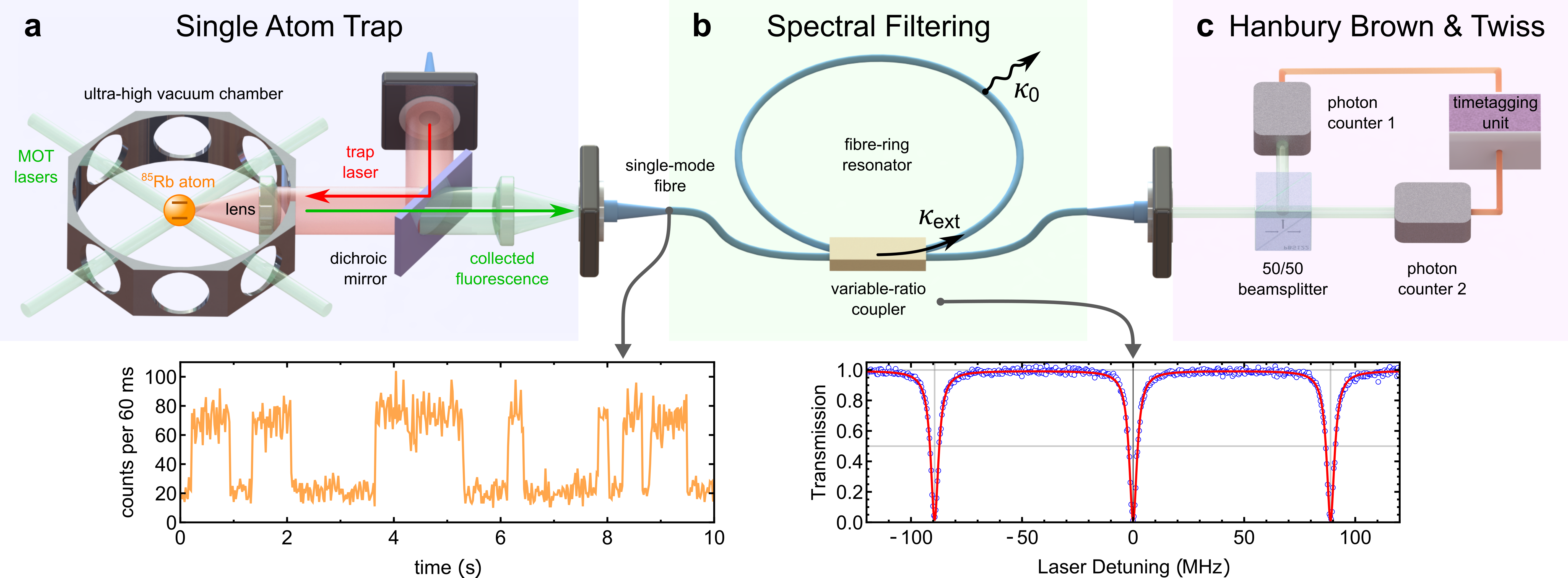}
\caption{\label{ExperimentalSetUp}\textbf{Experimental set-up.} (a) A single $^{85}$Rb atom is loaded from a MOT into an optical dipole trap. Fluorescence photons from the atom are collected with a lens (NA $= 0.55$), separated from the trapping laser with a dichroic mirror, and coupled into a single-mode fibre. Inset: Measured time trace of the photon countrate indicating the presence or absence of a single atom inside the trap. (b) The collected fluorescence is fibre-guided to an optical notch-filter realised by means of a fibre-ring resonator. A variable-ratio coupler allows for control over the coupling rate, $\kappa_{ext}$, into the resonator, which has an intrinsic loss rate of $\kappa_0$. A piezo-electric fibre stretcher (not displayed) is used for stabilisation of the resonator to the MOT laser frequency $\omega_L$. A fibre-polarisation controller (not displayed) is used to compensate for birefringence in order to realise a non-polarisation-selective filter operation. Inset: Transmission spectrum for the critically-coupled filter resonator, i.e., when $\kappa_{ext} = \kappa_0$. This spectrum shows a full-width half-maximum linewidth of $4.4$ MHz with a free spectral range of $89.1$ MHz. (c) The filtered fluorescence light is sent to a Hanbury Brown \& Twiss set-up consisting of a pair of single photon counters behind a $50/50$ beamsplitter. Photon arrival times are recorded by a timetagging unit, from which we measure the second-order correlation function, $g^{(2)}(\tau)$, of the filtered light.}
\end{figure*}

\par Here, we present an experiment that clearly demonstrates the validity of this interference picture by, via spectral filtering, selectively collecting only the incoherently scattered two-photon component. For this, we make use of an adjustable, narrow-band optical notch filter, which reduces the amplitude of the coherently scattered light. Subsequently, we measure the second order correlation function, $g^{(2)}(\tau)$, of the residually transmitted light. By tuning the relative magnitude of the two scattered components, we observe an evolution from a clear photon antibunching of $g^{(2)}(0) = 0.43 \pm 0.02$ without filtering, to a strong photon bunching of up to $g^{(2)}(0) = 7.65 \pm 0.81$ when maximally rejecting the coherently scattered light. Our observation indicates that, counterintuitively, the incoherently scattered light only consists of simultaneously scattered photons.
\par In our experiment, we prepare a single $^{85}$Rb atom in an optical dipole trap that is loaded from a magneto optical trap (MOT). The MOT lasers, with frequency $\omega_L$, are red-detuned with respect to the Stark-shifted atomic resonance by $\Delta/2\pi = -57.9 \pm 3.7$ MHz. In this setting, the atom scatters photons from the MOT laser beams into free space. The quantum state of this fluorescence light can be separated into a coherent state, $\lvert\alpha\rangle$, and an incoherently scattered component, $\lvert\phi\rangle$, whereby the latter originates from the saturable nature of the emitter. We weakly drive the atom with a saturation parameter $S = 0.025 \pm 0.004$. In this low-saturation regime, the incoherently scattered part consists solely of energy-time entangled photon pairs \cite{Shen2007,Mahmoodian2018,prasad20,hinney21}, with a probability amplitude for finding two photons within a time delay $\tau$ given by
\begin{eqnarray}
\phi^{(2)}(\tau)=-\frac{n_\textrm{coh}}{2}e^{-(\gamma-i\Delta)\lvert\tau\rvert}.
\end{eqnarray}
The corresponding two-photon amplitude for the coherent component is given by the time-independent value
\begin{eqnarray}
\alpha^{(2)}=\frac{n_\textrm{coh}}{2},
\end{eqnarray}
see Methods for details. Here, ${n_\textrm{coh}=\gamma S/(S+1)^2}$ is the photon scattering rate of the coherent component and $\gamma$ is the amplitude decay rate of the atomic dipole. The total scattered field is given by the sum of the two components $\lvert\psi\rangle=\lvert\alpha\rangle+\lvert\phi\rangle$ and for its two-photon amplitude one obtains 
\begin{equation}\label{Psi2}
\begin{split}
\psi^{(2)}(\tau)&=\alpha^{(2)}+\phi^{(2)}(\tau)\\&=\frac{n_\textrm{coh}}{2}\left(1-e^{-(\gamma-i\Delta)\lvert\tau\rvert}\right).
\end{split}
\end{equation}
Importantly, at zero time delay ($\tau=0$), the coherently and incoherently scattered components have equal amplitude but are $\pi$ out of phase. This results in perfect destructive interference such that the light never contains two simultaneously scattered photons. Despite this fact, taken individually, both the coherent and incoherent components have a non-zero probability for containing two simultaneous photons.
\par Now, the two components exhibit different spectra, which for the coherently scattered component is given by a delta function at $\omega_L$ while for the incoherently scattered component, at low saturation, the spectrum is approximately given by a pair of Lorentzian lines, each of width $\gamma$, which are separated by $\pm\Delta$ from $\omega_L$ \cite{Steck07}, see Fig. \ref{Experiment_concept}. Their distinct spectra allow us to separate the components by applying a frequency notch-filter centered at the driving laser frequency. For this, we first collect the atomic fluorescence with a high numerical-aperture objective before coupling it into a single-mode fibre which guides the light to the filter, see Fig. \ref{ExperimentalSetUp}. The latter is realised by means of a fibre-ring resonator, where we use a low-loss variable ratio fibre beamsplitter to control the coupling rate into the resonator, $\kappa_{ext}$, and in turn the on-resonance transmission past the resonator, see rightmost inset of Fig. \ref{ExperimentalSetUp}. The resonator is stabilised to resonance with the driving laser and has an adjustable linewidth between $1...15$ MHz, which is much smaller than the separation of the coherent and incoherent components of $\lvert\Delta\rvert$. As a consequence, the filter reduces the coherent component by the complex valued field transmission factor, $t_F(\kappa_{ext})$, but does not significantly alter the incoherent component, see Methods. To measure the modified photon statistics after filtering, we send the light to a Hanbury Brown \& Twiss set-up consisting of a $50/50$ beamsplitter and a single photon counter in each of its outputs \cite{HBT56} and record the second-order correlation function, $g^{(2)}(\tau,\kappa_{ext})\propto|t_F(\kappa_{ext})^2\alpha^{(2)}-\phi^{(2)}(\tau)|^2$, for different settings of the coupling parameter $\kappa_{ext}$.

%width=0.446\textwidth
\begin{figure}[ht]
\centering
    \includegraphics[height=0.824775\textheight,keepaspectratio]{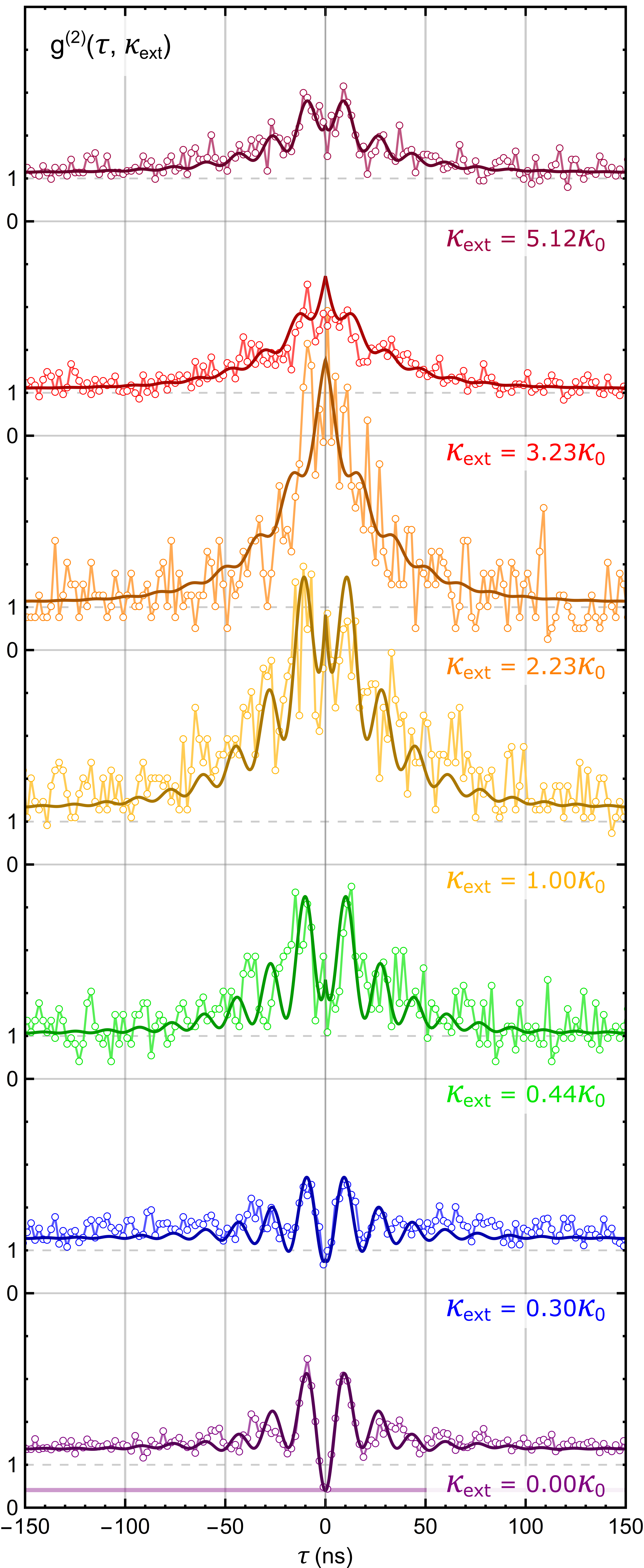}
\caption{\label{Waterfall} \textbf{Measured second order correlation functions $\bm{g^{(2)}(\tau)}$ for different filter settings $\bm{\kappa_{ext}}$.} Each data set has the same vertical scale and is offset by 5. The solid lines are fits of our theoretical model to the data points (empty circles), with the temperature of the trapped atom and a residual resonator--laser detuning as fit parameters, see Methods. For the unfiltered data set ($\kappa_{ext} = 0$), the purple horizontal line indicates the noise floor due to background counts.}
\end{figure}

\par Figure \ref{Waterfall} showcases the second order correlation functions as $\kappa_{ext}$ is increased. For $\kappa_{ext} = 0$, the filter resonator has no effect on the collected atomic fluorescence. The resulting measured second order correlation function (purple) shows $g^{(2)}(0) = 0.43 \pm 0.02$ which is compatible with perfect photon antibunching when considering the background countrates of our photon counters. This illustrates the expected behaviour that, in the absence of spectral selection, the light scattered by a single atom never contains more than one photon at the same time and place. For $\lvert\tau\rvert > 0$, we observe a damped oscillatory behaviour, originating from the driven atom undergoing Rabi oscillations at an effective frequency of $\Omega_{\textrm{eff}} \approx \Delta$, which are dampened on a timescale of the excited state lifetime, $1/2\gamma$. This oscillatory behaviour can also be seen as to originate from interference between the coherently and incoherently scattered two-photon components, cf. Eq. (\ref{Psi2}).
\par When increasing $\kappa_{ext}$, the power transmission, $\lvert t_F(\kappa_{ext})\rvert^2$, of the coherent component through the filter is reduced. Consequently, we upset the delicate balance of the two components, resulting in less destructive interference and hence an increasing $g^{(2)}(\tau = 0)$ (blue and green data sets). Around $\kappa_{ext} \approx \kappa_0$, the coherent component is mostly removed. Consequently, the second order correlation function resembles a double-exponential envelope centred at $\tau = 0$, featuring strong photon bunching of up to $g^{(2)}(\tau = 0) = 7.65 \pm 0.81$ (yellow and orange data sets). When further increasing $\kappa_{ext}$, the transmission, $t_F(\kappa_{ext})$, of the coherent component will increase again and the bunching decreases (red and violet data sets). For very large values of $\kappa_{ext}$, antibunching would be recovered, although not observed here for technical reasons.

\begin{figure}[t]
\centering
\includegraphics[width=0.48\textwidth]{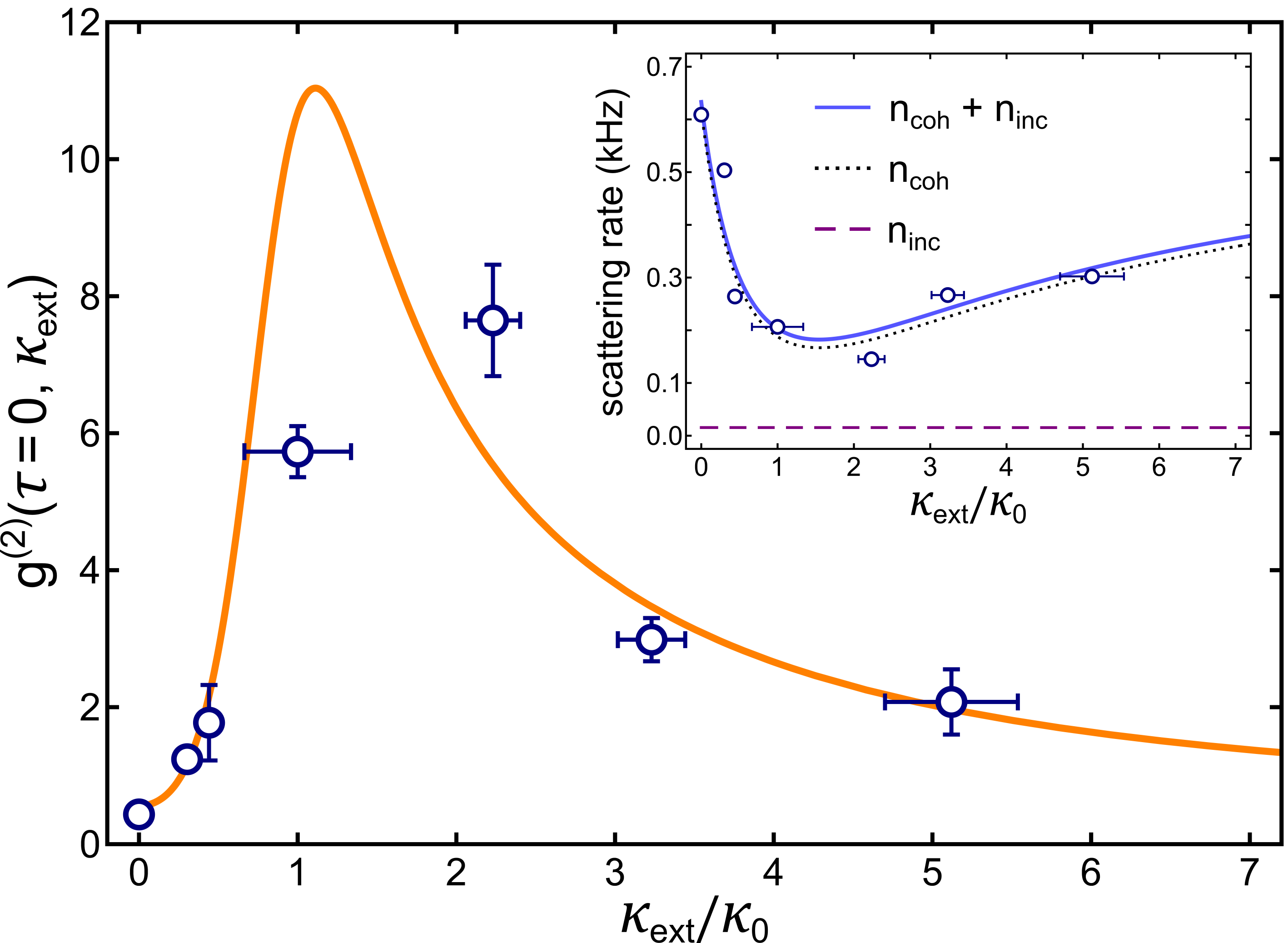}
\caption{\label{g20vsKext} \textbf{Correlations at zero time delay as a function of the filter setting.} The data points are obtained from the data shown in Figure \ref{Waterfall}, with the error bars indicating the $1\sigma$ uncertainty of $\kappa_{ext}$ and $g^{(2)}(0)$. The solid line is the theoretical prediction, see main text for details. Inset: Dark count corrected detected photon rate after filtering (data points), with the corresponding theory prediction (blue solid) which contains the single photon detection efficiency $\eta$ as only fit parameter. From the fit we obtain $\eta=0.136\pm0.002\%$. The individual contributions to the scattering rate are depicted by the black dotted (coherent) and purple dashed (incoherent) lines. The horizontal error bars are the same as in the main figure, while those in the vertical direction are not visible on the scale of the inset.}
\end{figure}

\par Figure \ref{g20vsKext} shows the evolution of the measured second-order correlation function at $\tau = 0$, as a function of $\kappa_{ext}$. For each setting, the value of $g^{(2)}(0,\kappa_{ext})$ is obtained by averaging the data shown in Fig. \ref{Waterfall} in a $3$ ns window centred around $\tau = 0$. The solid orange curve is the theoretical prediction of $g^{(2)}(0,\kappa_{ext})$, see Methods. We note that our model takes into account residual drifts of the filter resonance, which we obtain from fits to the data sets in Fig. \ref{Waterfall}, see Methods. As a consequence of these drifts, maximum suppression of the coherently scattered component is reached for the setting of $\kappa_{ext}$ slightly larger than $\kappa_0$. 
\par From the two data sets exhibiting the best suppression of the coherently scattered light ($\kappa_{ext} = 1.00\kappa_0$ and $\kappa_{ext} = 2.23\kappa_0$), we obtain a rate of detected photon pairs of $n^{(2)}_{\textrm{meas.}} = (4.91 \pm 0.47) \times 10^{-3}$ s$^{-1}$, see Methods. Given the pair-detection efficiency $\eta^2/2$ of our set-up, this corresponds to a total rate of photon pairs incoherently scattered by the atom of $n^{(2)}_{\textrm{total}} =  5.3 \pm 0.5$ kHz. This agrees well with the rate of incoherently scattered photon pairs, $n_{\textrm{inc}}/2 = 5.7 \pm 1.4$ kHz, that is expected for our saturation parameter, see Methods. This validates the picture that the incoherently scattered light consists of simultaneously scattered photon pairs.
\par To conclude, we return to the initial question on whether a single two-level atom will simultaneously scatter two photons. As is often the case in quantum mechanics, the answer is both yes and no, depending on which picture is considered. Indeed, the unmodified light field emitted by the atom never contains two simultaneous photons. Here, we have highlighted the picture where this is a consequence of the atom continuously scattering the incident two-photon component in two different ways, which interfere perfectly destructively. An alternative picture is to state that the atom only scatters photons one by one, and that the emitted stream of antibunched photons is transformed into bunched photons by the passive linear notch filter in our set-up. Surprisingly, while this seems at odds with standard intuition, the latter is also revealed to be a valid interpretation of our observations if one evokes a Hong-Ou-Mandel quantum interference effect \cite{HOM87} at the incoupling beam splitter, see Methods. Beyond these fundamental considerations, the demonstrated effect lends itself to realising spectrally narrowband photon pair sources that are inherently compatible with optical transitions in quantum emitters and highly sought after in optical quantum technologies.

\begin{acknowledgments}
We acknowledge funding by the Alexander von Humboldt Foundation in the framework of the Alexander von Humboldt Professorship endowed by the Federal Ministry of Education and Research, as well as the European Commission under the project DAALI (No.899275). X. H. acknowledges a Humboldt Research Fellowship by the Alexander von Humboldt Foundation. M. C. and M. S. acknowledge support by the European Commission (Marie Sk\l{}odowska-Curie IF Grant No. 101029304 and IF Grant No. 896957). \\
\end{acknowledgments}

\textit{Note added:} whilst preparing this manuscript, we became aware of related work by B. L. Ng et al. \cite{ng22}.

\bibliography{PhotonPair} 
\clearpage
\newpage
\begin{center}
    \large\textbf{\uppercase{METHODS}}
\end{center}
\preprint{APS/123-QED}

%\title{\MakeUppercase{METHODS}}
%\maketitle
%\onecolumngrid
\section{Theoretical Description}

\subsection{Wavefunction description of the scattered field}
In a standard picture, the atom-light interaction is described by optical Bloch equations, in which the atom is excited by a coherent laser beam and the scattered quantum field can be calculated from the time evolution of the atomic raising and lowering operators \cite{Steck07}. In the steady state, the scattered field exhibits two distinct components, usually referred to as coherently and incoherently scattered light, where their ratio depends on the saturation parameter of the excitation laser field
\begin{equation}
S=\frac{1}{2}\frac{(\Omega/\gamma)^2}{1+(\Delta/\gamma)^2}.
\end{equation}
Here, $\Delta=\omega_{L}-\omega_{0}$ is the laser-atom detuning with the laser frequency $\omega_L$ and the atomic resonance frequency $\omega_0$. Furthermore, $\Omega$ is the Rabi-frequency of the driving field and $\gamma$ is the amplitude decay rate of the excited state. The coherent and incoherent scattering rates of the atom are given by
\begin{align}
n_\textrm{coh}&=\gamma\frac{S}{(S+1)^2}\approx\gamma (S-2S^2)\label{ncoh}\\
\intertext{and}
n_\textrm{inc}&=\gamma \frac{S^2}{(S+1)^2}\approx \gamma S^2, \label{ninc}
\end{align}
respectively, where the right-hand-side expressions are the low saturation expansions up to second order in $S$, which we apply throughout the manuscript in order to obtain a simple theoretical description of the scattered light. In the following, we consider the case where we drive the atom with an excitation laser beam for a time duration $T$, where $T$ is chosen such that it is much larger than the atomic decay time, $T\gg1/\gamma$, in order to neglect transient features and small enough such that the mean photon number is much smaller than one, i.e. $T\ll1/\gamma S$. We note that in the low saturation regime, $S\ll1$, this is always possible. The coherently scattered light can then be represented as a coherent state
\begin{equation}
|\alpha\rangle\approx|0\rangle+\alpha|1\rangle+\frac{\alpha^2}{\sqrt{2}}|2\rangle    
\end{equation} 
with the mean photon number $\alpha^2=n_\textrm{coh}T$ where, without loss of generality, we assumed $\alpha$ to be real. In the time domain, in a frame rotating with the laser frequency, the two-photon component of the state can be written as
\begin{equation}
\frac{\alpha^2}{\sqrt{2}}|2\rangle=\iint\limits_{-T/2}^{T/2}  dt_1\, dt_2\,\underbrace{\frac{\alpha^2}{2T}}_{\alpha^{(2)}} a^\dagger_{t_1} a^\dagger_{t_2}|0\rangle
\end{equation}
where $a^\dagger_{t}$ ($a_{t}$) is the creation (annihilation) operator for a photon at time $t$ and $\alpha^{(2)}=n_\textrm{coh}/2$ is the amplitude of the two-photon wavefunction.
In the low saturation limit, the incoherently scattered light consists of energy-time entangled photon pairs \cite{Shen2007,Mahmoodian2018,prasad20,hinney21} and the state in the time domain can be written as
\begin{equation}
|\phi\rangle=\iint\limits_{-T/2}^{T/2} dt_1\, dt_2\, \phi^{(2)}(t_2-t_1)a^\dagger_{t_1} a^\dagger_{t_2}|0\rangle.
\end{equation}
Here the integrals extend over the time duration $T$ and the temporal envelope $\phi^{(2)}(t_2-t_1)$ is given by
\begin{equation}
\phi^{(2)}(t_2-t_1)=-\frac{n_\textrm{coh}}{2}e^{-(\gamma-i\Delta)|t_2-t_1|}.
\end{equation}
The total scattered field is thus given by
\begin{equation}
|\psi\rangle=|\alpha\rangle+|\phi\rangle,
\end{equation}
and for its two-photon component one obtains 
\begin{equation}
|\psi^{(2)}\rangle=\frac{n_\textrm{coh}}{2}\iint\limits_{-T/2}^{T/2} dt_1\, dt_2\,\big[1-e^{-(\gamma-i\Delta)|t_2-t_1|}\big] a^\dagger_{t_1} a^\dagger_{t_2}|0\rangle.\label{eq:psi2}
\end{equation}

\subsection{Modelling the photon statistics}
The spectral distributions of the two components differ from each other, where the spectrum of the coherent state $|\alpha\rangle$ is approximately given by a delta function at the laser frequency $\omega_L$, whereas the spectrum of the incoherently scattered component, at low saturation, is approximately given by a pair of Lorentzian lines centered at the frequencies $\omega_L\pm\Delta$. This frequency separation between the coherent and incoherent component allows us separate them by using a narrowband notch filter at the laser frequency, which in our experiment is realised as a fibre-ring resonator, see section \ref{FRR}. As the filter bandwidth is much smaller than the separation of the frequency components, we make the simplifying assumption that the filter acts only on the coherently scattered light and reduces its field amplitude by the complex-valued field transmission factor \cite{zeiger17}
\begin{equation}
t_F(\omega)=\frac{\kappa_{0}-\kappa_{\textrm{ext}}+i (\omega-\omega_\textrm{res})}{\kappa_{0}+\kappa_{\textrm{ext}}+i (\omega-\omega_\textrm{res})}.\label{eq:tF}
\end{equation}
Here, $\kappa_{\textrm{ext}}$ is the external coupling rate into the filter resonator, $\kappa_0$ the intrinsic loss rate of the filter resonator and $\omega_\textrm{res}$ its resonance frequency. After filtering, the wavefunction of the light is given by
\begin{equation}
|\psi_F\rangle=|t_F(\omega_L)\cdot\alpha\rangle+|\phi\rangle. \label{eq:filtered}
\end{equation}
In our experiment, we analyse the photon statistics of the filtered atomic fluorescence by measuring its second order correlation function 
\begin{eqnarray}
g^{(2)}(\tau)&=&\frac{\langle a^\dagger_ta^\dagger_{t+\tau}a_{t+\tau}a_t\rangle_t}{\langle a^\dagger_ta_t\rangle_t^2}\equiv\frac{G^{(2)}(\tau)}{n^2}
\end{eqnarray}
where $\langle...\rangle_t$ indicates the time-averaged expectation value and $\tau = t_2 - t_1$ is the temporal separation between the two photons. $G^{(2)}(\tau)\propto\eta^2|\psi_F^{(2)}|^2$ is the unormalised second order correlation function for which we obtain from Eq. (\ref{eq:filtered})
\begin{eqnarray}
G^{(2)}(\tau)&\approx&\gamma^2 S^2\eta^2 \cdot \Big|t_F(\omega_L)^2-e^{-(\gamma-i\Delta)|\tau|}\Big|^2.\label{eq:bigG}
\end{eqnarray}
The mean photon flux, $n$, after the filter is given by
\begin{eqnarray}
n&=& \gamma S \eta\cdot\left(|t_F(\omega_L)|^2(1-2S)+S\right)\nonumber \\
&=&\eta\left(|t_F(\omega_L)|^2 n_\textrm{coh}+n_\textrm{inc}\right).
\end{eqnarray} 
Here, we introduced the total single-photon collection and detection efficiency of our setup, $\eta$. In the experiment, in addition to the measured photon rate, we also have to take into account the total background countrate, $d$, that originates from stray light and detector dark counts. Including this background in the experimentally expected second order correlation function, we obtain
\begin{eqnarray}
g^{(2)}_\textrm{exp}(\tau)&\approx&\frac{G^{(2)}(\tau)-n^2}{(n+d)^2}+1. \label{g2exp}
\end{eqnarray}

\subsection{Description in the time domain}
The transformation of the stream of antibunched photons emitted by the atom into a stream of bunched photons after the notch filter can also be explained by evoking Hong-Ou-Mandel quantum interference \cite{HOM87}: To this end, we make use of the fact that the resonator has a slow temporal response to the light field with a characteristic time scale $(\kappa_{0}+\kappa_{\textrm{ext}})^{-1}$ that is much slower than the amplitude fluctuations in the atomic fluorescence which are on a timescale of $\approx\Delta^{-1}$. Consequently, the resonator averages over these fluctuations and, in the steady state, the light stored in the resonator can approximately be described by a coherent state with amplitude
\begin{equation}
\alpha_r =-\sqrt{n_\textrm{coh}} \frac{\sqrt{2\kappa_{\textrm{ext}}}}{i(\omega-\omega_\textrm{res})+\kappa_{\textrm{ext}}+\kappa_0}.
\end{equation}
To calculate the probability of a photon coincidence detection with time delay $\tau$ after the filter resonator, we have to consider the different ways in which such a detection event can occur. In this picture, there are two fields, the resonator field and the incident fluorescence, that are overlapped on the incoupling beamsplitter of the resonator. Two photons can either originate both from the original fluorescence light (amplitude: $\psi^{(2)}_f$), both from the resonator (amplitude: $\psi^{(2)}_r$), or one photon from each (amplitude: $\psi^{(2)}_{fr}$). Note that the incoupling beamsplitter is highly asymmetric with a transmission of approximately one for the incident fluorescence light. The three probability amplitudes are then given by
\begin{eqnarray}
\psi^{(2)}_f&=&\frac{n_\textrm{coh}}{2}\big(1-e^{-(\gamma-i\Delta)|\tau|}\big)\\
\psi^{(2)}_r&=&\frac{(-\alpha_r\sqrt{2\kappa_{\textrm{ext}}})^2}{2}=\frac{n_\textrm{coh}}{2} \big(t_F(\omega_L)-1\big)^2\\
\psi^{(2)}_{fr}&=&-\sqrt{n_\textrm{coh}}\cdot\sqrt{2\kappa_{\textrm{ext}}}\,\alpha_r=n_\textrm{coh}\big(t_F(\omega_L)-1\big).
\end{eqnarray}
The total two-photon detection probability amplitude is the coherent sum of the three components and we obtain  for the two photon field after the resonator
\begin{eqnarray}
\psi_{F}^{(2)}&=&\psi_f^{(2)}+\psi_r^{(2)}+\psi_fr^{(2)}\nonumber\\
&=&\frac{n_\textrm{coh}}{2}\big[t_F(\omega_L)^2-e^{-(\gamma-i\Delta)|\tau|}\big].
\end{eqnarray}
This is the same behaviour as in Eq. (\ref{eq:bigG}) which demonstrates the equivalence of the two pictures. 

\section{Experimental Methods}
\subsection{Trapping and Detecting Single Atoms}
We prepare a cloud of $^{85}$Rb atoms inside an ultra-high vacuum chamber using a magneto-optical trap (MOT). The MOT lasers, with frequency $\omega_L$, are red-detuned with respect to the unperturbed D$_2$ transition of $^{85}$Rb (transition frequency $\omega_0$) by $\delta_{MOT} = \omega_L - \omega_0 = - 2\pi \times 16.3~$MHz. The MOT cloud contains several million $^{85}$Rb atoms and is used as a reservoir of cold atoms for loading the optical dipole trap. To load a single atom, the MOT cloud is positioned at the focus region of an aspheric lens with a Numerical Aperature of $0.55$ (AS-AHL12-10, Asphericon), that is located inside the vacuum and has a focal length of $f = 10$ mm and a working distance of $w_d = 7.6$ mm. The optical dipole trap is generated by focusing the incident trap laser beam (wavelength: $\lambda_{\textrm{trap}} = 784.65$ nm) to a waist radius of $w_0 = 1.8 \pm 0.2$ µm inside the MOT cloud. For a laser power of $P_{\textrm{trap}} = 2.5$ mW, we obtain an optical trapping potential with a depth of $U_{\textrm{trap}}/k_B = 1.66$ mK, corresponding to trap frequencies of $\omega_r = 2\pi \times 96$ kHz and $\omega_z = 2\pi \times 17$ kHz in the radial and axial directions, respectively. The photon scattering rate of the trapping light field amounts to $0.86$ kHz.
\par
\begin{figure}[t]
\centering
   \includegraphics[width=0.4\textwidth]{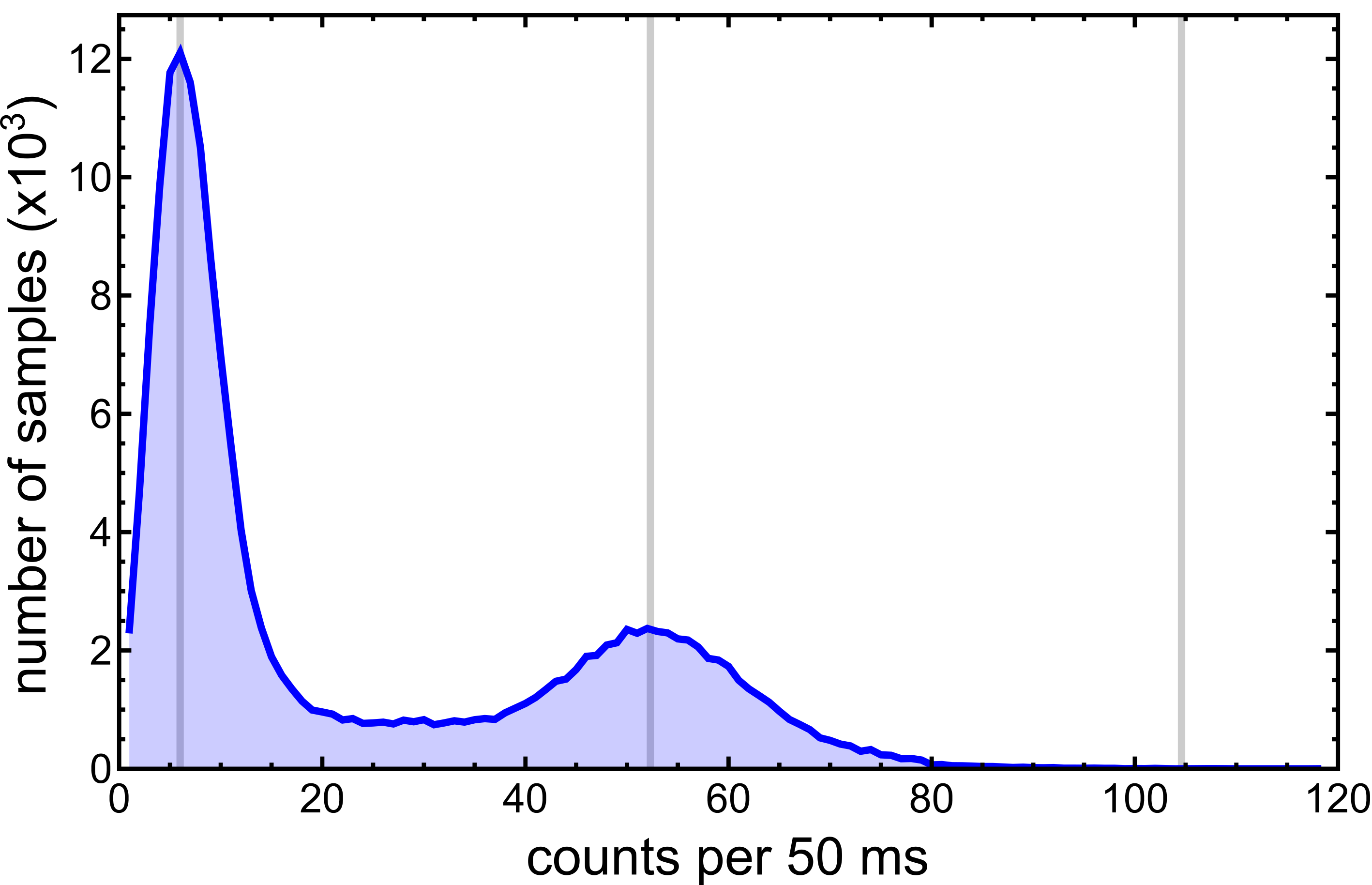}
\caption{\textbf{Histogram of the detected photon countrate originating from the trap volume.} Without spectral filtering, we observe an increase in the most likely countrate from a background of $120$ s$^{-1}$, to $1050$ s$^{-1}$ when one atom is trapped. The absence of higher countrates indicates the sub-poissonian occupation statistics due to the collisional blockade effect present in microscopic optical traps \cite{Schlosser2001}.}\label{fig:atomhist}
\end{figure}

 The presence of an atom in the trap is registered by an increase of the photon detection rate from $120$ s$^{-1}$ to $1050$ s$^{-1}$ for the case without spectral filtering, see Fig. \ref{fig:atomhist}. Here, we collect the fluorescence light with the same lens that we use for focussing the trap beam. Due to the microscopic trap volume, our trap operates in the collisional blockade regime \cite{Schlosser2001} such that, at most, a single atom is present inside the trapping volume at any time, see Fig. \ref{fig:atomhist}. The collected fluorescence light is separated from the trapping light with a bandpass dichroic mirror centered at $780$ nm that features a $3$-nm linewidth (LL01-785-25, Semrock). We then couple the light into a single-mode fibre which also acts as a spatial filter. The fibre-guided fluorescence is sent, via the fibre-ring resonator filter (see section \ref{FRR}), to a fibre-based Hanbury-Brown \& Twiss set-up consisting of two fibre-coupled single photon counting modules (SPCM) behind a $50/50$ fibre-optic coupler. The two SPCMs are connected to an FPGA-based timetagger which records the arrival times of each detected photon. 

\subsection{Spectral Filtering}\label{FRR}

\begin{figure}[t]
\centering
   \includegraphics[width=0.4\textwidth]{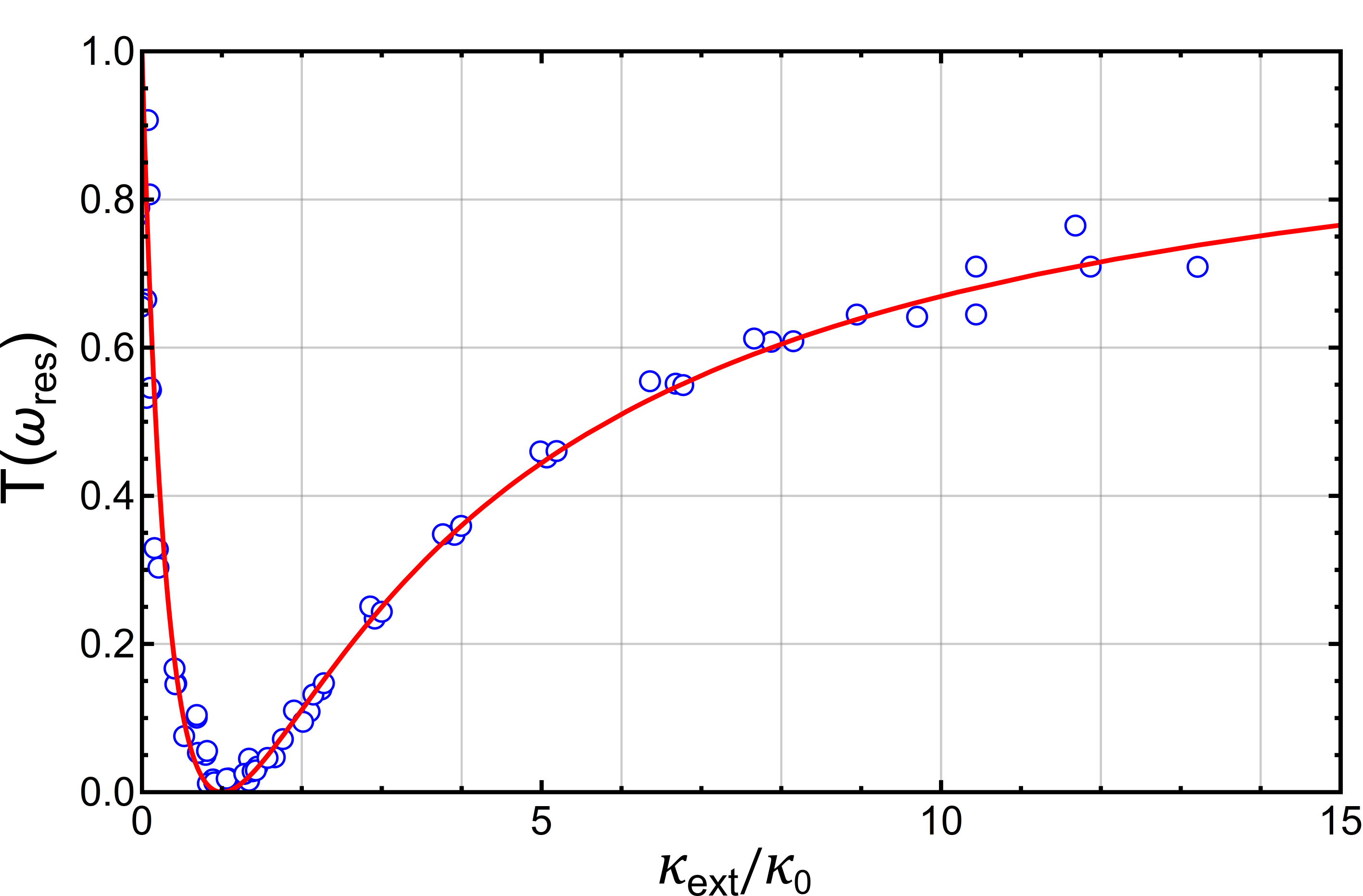}
\caption{\textbf{Characterisation of the spectral filter.} Measured on-resonance power transmission $|t_F(\omega_{\textrm{res}})|^2$ as a function of $\kappa_{\textrm{ext}}$ (blue data points). The solid red line is a fit according to Eq. \eqref{eq:tF}, yielding an intrinsic resonator loss rate of $\kappa_0/2\pi = 1.08\pm0.02~$MHz.}\label{fig:characterisation}
\end{figure}

Our fibre-based optical notch filter consists of a variable ratio coupler (F-CPL-830-N-FA, Newport) which allows setting of the coupling rate, $\kappa_{\textrm{ext}}$, into the resonator. A polarisation controller is included in the fibre-ring section, such that the two polarisation eigenmodes of the resonator are degenerate and the resonator acts as a polarisation-independent filter. This is important for consistent spectral suppression of the collected atomic fluorescence. A section of the fibre-ring is glued to a piezo-electric stack, which allows us to strain-tune the resonance frequency, $\omega_{\textrm{res}}$. The resonator is passively stabilized by placing its whole set-up inside a thermally and acoustically insulating box, thereby isolating the set-up from external environmental fluctuations. To compensate residual, slow drifts of its resonance frequency, every 5 s, we launch a stabilisation light field at frequency $\omega_L$ into the resonator set-up, scan the resonator length, and set it such that the point of minimum transmission is reached. The resonator has a geometrical length of $l = 2.25 \pm 0.05$ m, yielding a free spectral range of $\nu_{FSR} = 89.0 \pm 0.5$ MHz.
Figure \ref{fig:characterisation} shows the on-resonance transmission of the resonator for different coupling rates $\kappa_{\textrm{ext}}$. From a fit to the data, we obtain an intrinsic resonator loss rate of $\kappa_0 = 2\pi \times 1.08 \pm 0.02~$MHz, which results in an unloaded resonator finesse of $F = \pi\nu_{FSR}/\kappa_0 = 41.2\pm0.8$.

\begin{figure}[b]
\centering
\includegraphics[width= 0.43\textwidth]{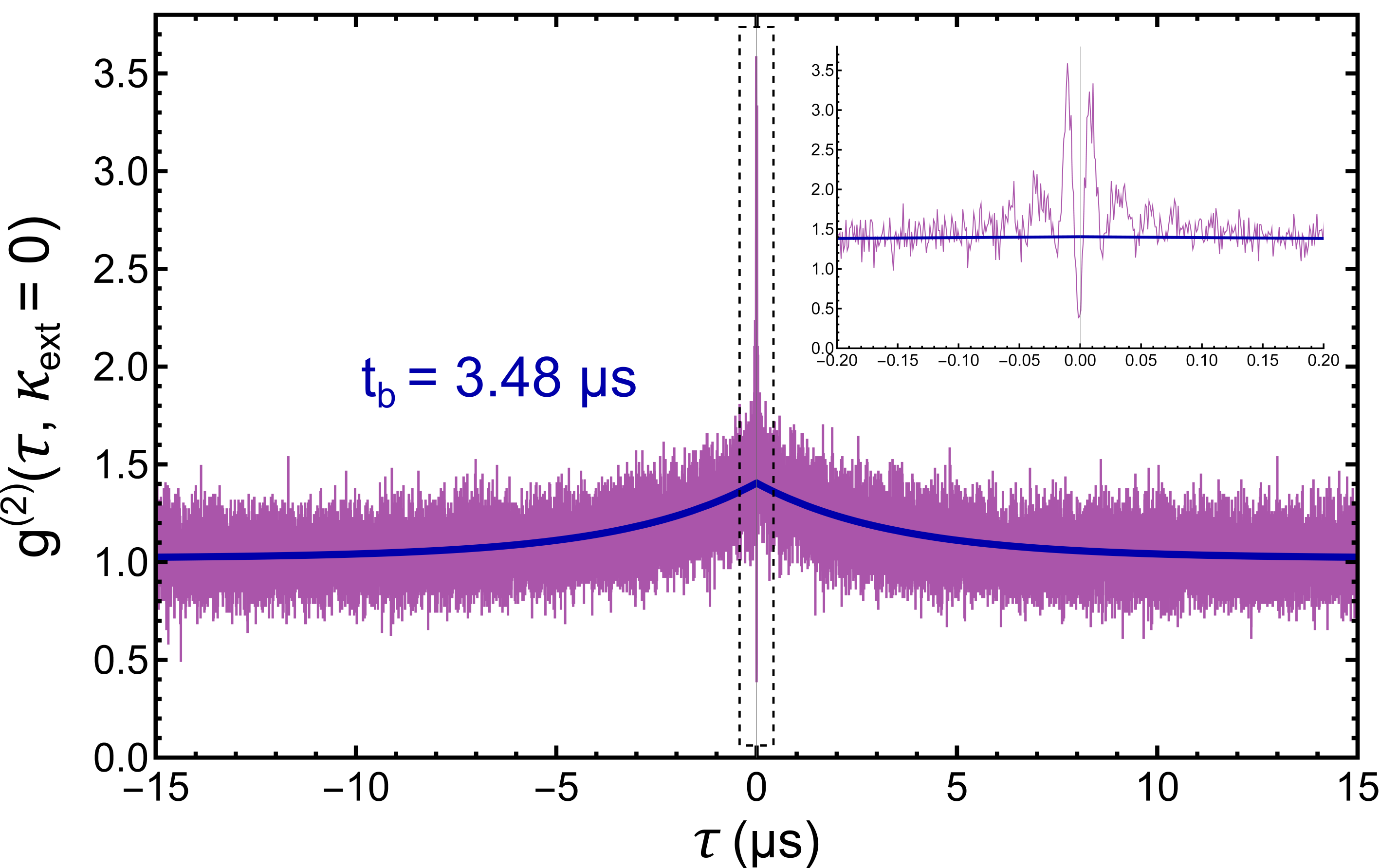}
\caption{\textbf{Long-term behaviour of the second order correlation function.} A weak bunching effect arising from diffusive atomic motion is apparent. It exhibits a decay time of $t_b = 3.48$ µs for the data set with $\kappa_{\textrm{ext}} = 0$, see main text. Inset: zoom on the area indicated by the dashed line. }\label{fig:technicalbunching}
\end{figure}

\subsection{Analysis of measured correlation functions}\label{analysis}
In order to fit the theory to the measured correlation functions in Fig. 3 we perform the following procedure.

First, the measured correlation functions exhibit a weak bunching effect on a µs-timescale, which originates from the diffusive motion of the trapped atoms through the complex intensity and polarization pattern of the MOT laser field \cite{Gomer98}, see Fig. \ref{fig:technicalbunching}.
To account for this, we fit the function $1 + A e^{\lvert\tau\rvert/t_b}$ to the data for large time delays $\lvert\tau\rvert\gg1/2\gamma$, from which we obtain typical values of $A = 1.38$ and ${t_b = 3.48}$ µs for the measurement without filtering ($\kappa_{\textrm{ext}}=0$). For each $g^{(2)}$ measurement, our model is then adjusted such that the value $1+A$ serves as the new baseline.
Second, the finite temperature of the atom in the dipole trap gives rise to a temperature-dependent distribution of atomic positions in the trap and, consequently, of the AC-Stark shifts. To account for this, we assume a thermal position distribution and fit the second order correlation function of the unfiltered case ($\kappa_{\textrm{ext}}=0$) with the atom temperature and the maximum AC-Stark shift as free fit parameters. The latter takes into account that the atom experiences a significant tensor light shift which results in a Zeeman state-dependent transition frequency.
From the fit, we obtain a temperature of $144\pm47$ µK and a mean detuning of the atom from the MOT lasers of $\Delta/2\pi = - 57.9 \pm 3.7~$MHz. These values are used to fit the data obtained for all other settings of the filter resonator.

Finally, for the measurements that incorporate the filter resonator, an additional experimental uncertainty occurs due to drifts of the narrow resonator absorption line with respect to the MOT laser frequency. To account for these residual resonator drifts in our model, we assume that the filter exhibits a Gaussian probability distribution of resonance frequencies with a width of $\sigma$ centered around an average resonator-laser detuning $\delta = \omega_{\textrm{res}} - \omega_L$. Using this model, we fit all measured correlation functions in Fig. 3 with $k_{\textrm{ext}}\neq0$. From the results we obtain an average laser-resonator detuning of $\delta = -0.26\pm0.58~$MHz with a width of $\sigma=1.92\pm0.36~$MHz.

\subsection{Detection efficiency}

From the atomic temperature obtained from the previous section, we calculate the distribution of saturation parameters $S$ from which we obtain the expected scattering rates of the incoherent and coherent components. Together with the laser-filter detuning we can then calculate the expected photon detection rate. We fit this model with the total photon detection efficiency, $\eta$, as the only free parameter to the data shown in the inset of Fig. 4. The fit yields $\eta=0.136\pm0.002~\%$, which agrees with the value expected from our collection efficiency of $\sim 1.3~\%$, together with the propagation losses through the filter setup and the limited detector efficiency of $\sim 0.55$.

\subsection{Correlation function at zero time delay vs. filter setting.}

To evaluate the theory prediction of the zero-time-delay value of the second order correlation as a function of $\kappa_{\textrm{ext}}$ shown in Fig. 4, we use the approximate expression for $g^{(2)}(\tau)$ given in Eq. (\ref{g2exp}). Here, we use the temperature-induced distribution of $S$, the residual laser-resonator detunings and the photon detection efficiency $\eta$, which we obtained from previous measurements, as well as the mean detector background rates as parameters.

\subsection{Photon Pair Rate}
To determine the number of detected photon pairs, we start with the histogram of measured coincidences and subtract the base level found for $\lvert\tau\rvert\gg1/2\gamma$, which stems from accidental coincidences. By summing up all remaining coincidences within a time interval of $\pm 100$ ns around $\tau = 0$ and dividing by the effective measurement time, we obtain the rate of detected photon pairs. Given our photon-pair detection efficiency $\eta^2/2$, we then determine the total rate of photon pairs scattered by the atom, $2n^{(2)}_{\textrm{meas.}}/\eta^2$. For the case of strong suppression of the coherent component, i.e. $t_F \approx 0$, these detected pairs all originate from the incoherently scattered light, thus the rate of scattered photon pairs should be given by $n_\textrm{inc}/2$, see Eq. \eqref{ninc}.

\end{document}